\documentclass[twocolumn,showpacs,amssymb,aps,nofootinbib,floatfix,superscriptaddress]{revtex4-1}

\usepackage{epsfig}

\begin{document} 
\hbadness=10000

\title{Mass hierarchy in identified particle distributions in proton-lead collisions}

\author{Piotr Bo\.zek}
\email{Piotr.Bozek@ifj.edu.pl}
\affiliation{AGH University of Science and Technology,
Faculty of Physics and Applied Computer Science, al. Mickiewicza 30, 30-059 Krakow, Poland}
\affiliation{The H. Niewodnicza\'nski Institute of Nuclear Physics PAN, 31-342 Krak\'ow, Poland}

\author{Wojciech Broniowski}
\email{Wojciech.Broniowski@ifj.edu.pl}
\affiliation{The H. Niewodnicza\'nski Institute of Nuclear Physics PAN, 31-342 Krak\'ow, Poland}
\affiliation{Institut de Physique Th\'eorique, CNRS/URA 2306, F-91191 Gif-sur-Yvette, France}
\affiliation{Institute of Physics, Jan Kochanowski University, 25-406 Kielce, Poland}

\author{Giorgio Torrieri}
\email{torrieri@fias.uni-frankfurt.de}
\affiliation{FIAS, J.W. Goethe Universit\"at, Frankfurt A.M., Germany}
\affiliation{Pupin Physics Laboratory, Columbia University, 538 West 120$^{th}$ Street NY 10027, USA}

\date{ver. 2, 31 August 2013}  

\begin{abstract}
We study the mass dependence for identified-particle average transverse
momentum and harmonic flow coefficients
in proton-lead (p-Pb) collisions, recently measured at the LHC.  
The collective mechanism in the p-Pb system predicts a specific mass ordering 
in these observables: the growth of the average transverse momentum with the
particle mass and a mass splitting of the elliptic flow coefficient,
i.e., smaller differential elliptic flow of protons than pions for $p_T<2$~GeV. 
This provides an opportunity 
to distinguish between the collective scenario and the mechanism based on the initial gluon dynamics 
in the evolution of the p-Pb system.
\end{abstract}


\maketitle

In this Letter we analyze the mass hierarchy in proton-lead (p-Pb) collisions for average transverse
momentum, harmonic flow coefficients, and the ``ridge'' correlations
 \cite{CMS:2012qk,Chatrchyan:2013nka,Abelev:2012ola,Aad:2012gla,Aad:2013fja}, recently 
measured for identified particles at the LHC~\cite{Chatrchyan:2013eya,ABELEV:2013wsa,Abelev:2013bla}.
We show that the flow generated with hydrodynamics is capable of uniformly explaining these data for the most central p-Pb 
collisions, where collective effects are expected to be most important.

The fact that the (identified-particle) average transverse momentum $\langle p_T \rangle$ in  p-Pb collisions
 cannot be explained by a 
superimposition of p-p production was recently brought up by Bzdak and Skokov~\cite{Bzdak:2013lva}, 
thus on general grounds hinting collectivity, or strong departures from superposition in the p-Pb system.
We present below quantitative estimates showing that collective flow effects explain the observed
 $\langle p_T \rangle$ for identified particles.
As arguments based on saturation and on geometric scaling~\cite{McLerran:2013oju,Rezaeian:2013woa} lead to
similar phenomena, other observables are needed for the verification, in particular the particle-identified harmonic flow.   
The elliptic and triangular flow coefficients \cite{Ollitrault:1992bk,Alver:2010gr} in relativistic nuclear collisions (A-A)
arise as a result of the collective expansion of an azimuthally  asymmetric fireball.
In high multiplicity  A-A collisions, the flow asymmetry is routinely analyzed in terms of
the harmonic flow coefficients $v_n$,
\begin{equation}
\label{vndef}
\frac{dN}{d^3 p} = \frac{dN}{2\pi p_T dp_T d\eta} \left[1\!+\!2 \sum_{n=1}^\infty \!v_n(p_T,\eta) \cos\left(n(\phi-\Psi_n) \right)  \right].
\end{equation}

In proton-proton (p-p) and p-Pb  reactions at the LHC energies, the  two-particles distributions in relative azimuthal angle 
 $\Delta \phi$ and 
relative pseudorapidity $\Delta \eta$ have
been  analyzed \cite{Khachatryan:2010gv,CMS:2012qk,Abelev:2012ola,Aad:2012gla}.
The observation  of a sizable same-side  ($\Delta \phi\simeq 0$) and a away-side ridge ($\Delta \phi\simeq \pi$) 
in the highest multiplicity p-Pb events
resembles the effect noticed previously in the A-A case, where
the same-side ridge  and the broader away-side ridge 
occur as a result of azimuthally asymmetric collective expansion of the formed fireball~\cite{Takahashi:2009na}. 
The two-particle correlation function, including the ridges, can be successfully decomposed in Fourier components involving the squares of the 
harmonic flow coefficients, $v_n^2$. The same 
coefficients (up to the non-flow effects and flow fluctuations) are obtained from the flow analysis w.r.t. the event plane 
angles $\Psi_n$, or with cumulants~\cite{Luzum:2010sp}.
The coefficients $v_n$  reflect the structure of Fourier components of the initial transverse energy density, 
whose eccentricity coefficients are determined by the geometry of the collision and the fluctuations of 
the  initial  density in the transverse plane.

Hydrodynamic expansion of the small fireball formed in p-Pb collisions 
 generates relatively large elliptic and triangular flow 
\cite{Bozek:2011if}, and  may as well be behind  the  origin of the same-side ridge observed in p-Pb experiments at the LHC~\cite{Bozek:2012gr}.
The direct measurement of the elliptic flow, and especially of the triangular 
component $v_3$, strongly suggests a collective
origin of the effect \cite{Chatrchyan:2013nka,Aad:2013fja,ABELEV:2013wsa,Bozek:2011if,Bozek:2012gr,Bozek:2013df,Bozek:2013uha,Bzdak:2013zma,Qin:2013bha}.

In another class of models, the ridge is generated by 
local partonic dynamics~\cite{Kovner:2010xk,Dusling:2012iga},
which in the saturated regime leads to long range correlations in rapidity between the produced particles~\cite{Dusling:2009ni}.
The away- and same-side ridges are generated by the interplay of the ladder and interference diagrams~\cite{Dusling:2012iga,Dusling:2013oia}. 
The elliptic flow component in the two-particle correlations  can be explained in both scenarios. 
The existence of the two alternative scenarios of the dynamics in the p-Pb collisions 
which explain the observed ridge correlations calls for the evaluation
of additional experimental observables, able to disentangle different sources of two- or 
many-particle correlations. 

In this Letter we discuss how a further insight into the mechanism of particle production can be gained from the analysis of spectra and 
flow correlations for identified particles. In particular, we present the results for the average transverse momentum and the flow coefficients for  identified particles, and compare them 
to  very recent experimental results~\cite{ABELEV:2013wsa,Abelev:2013bla}.
It has been well known from the experience in the A-A studies that the flow generates a mass hierarchy in the $p_T$ spectra, where the transverse 
motion pushes heavier hadrons to higher
momenta~\cite{Schnedermann:1993ws}. 
The origin of the effect is very simple.
When an expanding hydrodynamic fluid freezes, it emits particles according to the Frye-Cooper~\cite{Cooper:1974mv} formula 
\begin{equation}
\frac{dN}{d^3 p} = \int_\Sigma dS(x,p),
\end{equation}
where the emission function is integrated over some freeze-out hypersurface. Explicitly, 
the ``boosted'' source element is~\cite{Cooper:1974mv}
\begin{equation}
\label{emission}
dS(x,p)= d \Sigma_\mu p^\mu f\left (-\frac{p_\mu u^\mu(x)}{T} \right),  
\label{Cooper:1974mv}
\end{equation}
with $f$ denoting the statistical distribution function at the freeze-out temperature $T$, 
 and $u$ is the flow four-velocity. 
The statistical hadronization includes also the important 
resonance contributions, but the qualitative aspects remain simple:
the momenta of heavier particles are affected more strongly by the collective flow; in 
particular, due to expansion, the protons will on the average acquire higher momentum than kaons, 
and those, in turn, higher than pions.

All results shown in this paper follow from the three-stage approach 
described in detail in~\cite{Bozek:2011if,Bozek:2013uha}. The fluctuating initial state is 
obtained with the Glauber simulations~\cite{Broniowski:2007nz}, where the initial 
density is constructed by placing a smeared source in the center of mass of the colliding proton and the 
participating nucleon (the {\em compact} source prescription of Refs.~\cite{Bzdak:2013zma,Bozek:2013uha}). 
We stress that the longitudinal elongation of the initial fireball in space-time rapidity is an assumption of the model 
that reproduces the observed pseudorapidity densities and leads to long range correlations in relative pseudorapidity of two particles. 
The subsequent event-by-event hydro simulations are for the 3+1~dimensional
 viscous dynamics, 
with the shear viscosity $\eta/s=0.08$ and the hydro ignition time of $\tau=0.6$~fm/c. 
The statistical hadronization~\cite{Chojnacki:2011hb} is carried out  
at the constant freeze-out temperature $T_f=150$~MeV. 

Some explanation of the centrality selection is in place. 
Exactly the same cuts in centrality as used by the experimental groups cannot be applied in
model calculations due to limited statistics. A cut in the initial entropy of the source, 
which is the prescription we adopt,  is an approximation of the  experimental
procedure. This method works for global average flow observables, especially in the range where their centrality 
dependence is mild, but cannot be applied to  very 
specific cases, such as the 
ultra-high multiplicity cuts of the CMS Collaboration, or used
to estimate the nuclear attenuation factors.
In the version of the Glauber Model used in this paper, for simplicity, the initial entropy is proportional to the number of participants 
and we define centrality classes via the number
of the participant nucleons in the Glauber Monte Carlo event~\cite{Bozek:2011if}. 
We note that using alternative scenarios~\cite{Bozek:2013uha} for the initial entropy does not affect the centrality dependence 
of the bulk observables studied in this paper.

\begin{figure}[t]
\epsfig{width=0.53\textwidth,figure=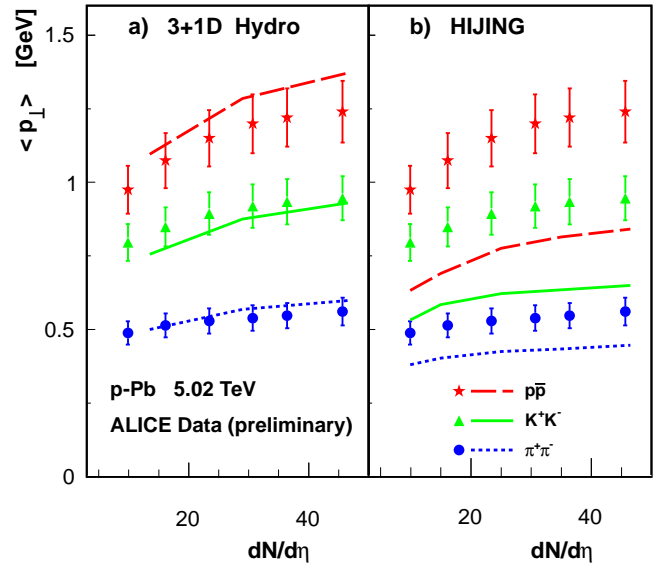}
\vspace{-15mm}
\caption{\label{figpt} Mean transverse momentum of identified particles as a function of the charged particle density
 in the p-Pb collisions, following from hydrodynamics~(a) and HIJING~2.1~(b).
The lines show the model calculations, while the data points come from Ref.~\cite{Abelev:2013bla}.
}
\label{fig01}
\end{figure}

Figure~\ref{figpt} shows the  mean transverse momenta of identified particles  in  two different approaches:  Hydro with the Cooper-Frye freeze-out (the model 
and its parameters are described in detail in~\cite{Bozek:2011if}), and the HIJING~ model \cite{Deng:2010mv}.
We note a large mass hierarchy in the hydro case, in agreement with the experiment, and supporting the collective scenario of the 
evolution. The hydrodynamic calculations are done for a range of centralities
$0-60$\%.
While the scenario assuming the collective expansion of the source
is justified only in the most central p-Pb collisions, our calculation reproduces 
the observed mass hierarchy and the multiplicity dependence of $\langle p_\perp \rangle$ for all centralities.
The HIJING model has no collective flow and cannot reproduce the measured mass 
splitting in the average transverse momentum. The fact that the superposition models do 
not reproduce the p-Pb data is visible  when using  the experimental data for $\langle p_\perp \rangle$ in p-p \cite{Bzdak:2013lva}. 
This also means that the color reconnection mechanism  which reproduces the $\langle p_\perp \rangle$
of the identified particles and their multiplicity dependence in p-p interactions 
\cite{Ortiz:2013yxa} would not explain the the p-Pb  data without additional collective flow or coherence effects. 

\begin{figure}[t]
\epsfig{width=0.49\textwidth,figure=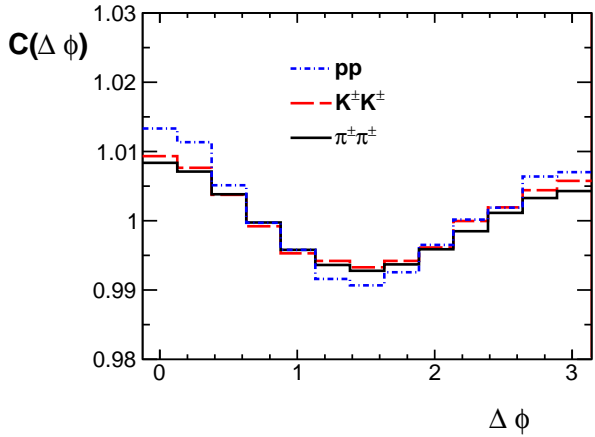}
\caption{
The correlation function $C(\Delta \phi)$ in the relative azimuth $\Delta\phi$ for identified particle pairs. The rapidity separation is $|\Delta\eta|>2$ and
the same-sign particles are used to minimize the non-flow effects. The momentum range is the same for all species: $0.3 < \langle p_T \rangle < 3$~GeV. 
Centrality is defined as $N_{part}\ge 18$, corresponding to $0-3\%$.
\label{figcorr}}
\end{figure}

\begin{figure}[tb]
\epsfig{width=0.53\textwidth,figure=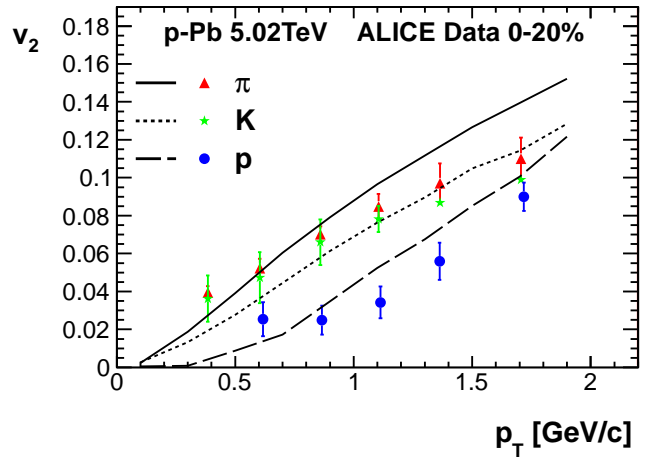}
\vspace{-5mm}
\caption{\label{figvn} $v_2\{2\}$ for pions, kaons and protons in p-Pb collisions calculated with the hydrodynamic model, 
as a function of the transverse momentum. The data come from Ref.~\cite{ABELEV:2013wsa}.}
\end{figure}

\begin{figure}[tb]
\epsfig{width=0.53\textwidth,figure=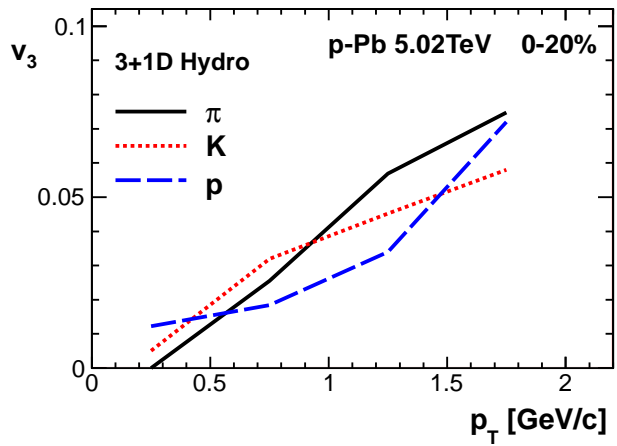}
\vspace{-5mm}
\caption{\label{figv3} Predictions for $v_3\{2\}$ for pions, kaons and protons in p-Pb collisions calculated with the hydrodynamic model, 
as a function of the transverse momentum.}
\end{figure}

In Fig.~\ref{figcorr} we show our model result for the identified correlations in the azimuthal angle, 
defined as 
\begin{eqnarray}
&& C(\Delta \phi,\Delta \eta)= \frac{S(\Delta \phi,\Delta \eta)}{B(\Delta \phi,\Delta \eta)}, \nonumber \\
&& C(\Delta \phi)= \int_{|\Delta\eta|>2} C(\Delta \phi,\Delta \eta). \label{eq:Cdef}
\end{eqnarray}
where $S$ is the distribution of signal pairs and $B$ is the mixed-event background.
 Note that Eq.~(\ref{eq:Cdef}) describes  the correlation function between two identified particles, while the flow coefficients $v_n(p_\perp)$ 
in Figs.~\ref{figvn} and \ref{figv3} correspond to the correlation of an identified particle with an unidentified charged hadron
\cite{ABELEV:2013wsa}. 
We  clearly  note the two ridges (maxima at $\Delta\phi=0$ and $\Delta\phi=\pi$) with amplitudes similar for all 
the studied particle species 
(the last observation is true only for the particular $p_\perp$ range used). 
The calculated correlation function should be compared to two-particle correlations extracted from the experiment
with the non-flow component subtracted.

Correlations from the collective flow in 
of Fig.~\ref{figcorr} represent the prediction
of the hydrodynamic model for correlations between two identical particle species. Thus in small systems
such flow correlations are present, besides 
the non-flow correlations from jets, local charge, momentum
or energy conservation, which are also observed in two-hadron correlations 
in p-p interactions \cite{Janik:2012ya}. 

\begin{figure}[tb]
\epsfig{width=0.53\textwidth,figure=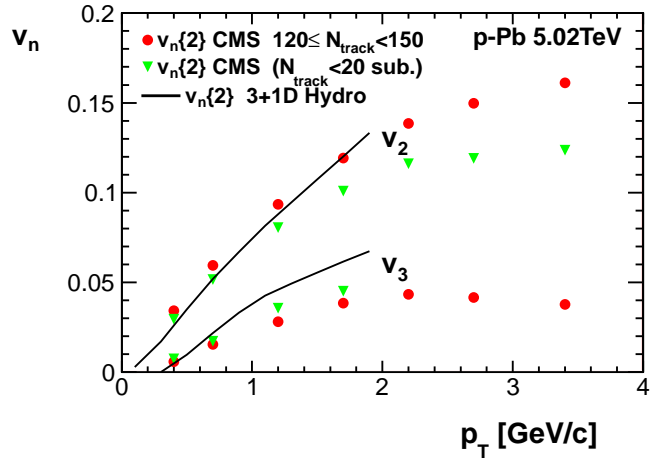}
\vspace{-5mm}
\caption{\label{v23pt} $v_2$ and $v_3$ for charged 
particles calculated with the hydrodynamic model. 
The data come from Ref.~\cite{Chatrchyan:2013nka}}
\end{figure}

The hydrodynamic model with Glauber initial conditions from the 
 {\em compact} source of Refs.~\cite{Bzdak:2013zma,Bozek:2013uha}
reproduces the measured elliptic and triangular flow of charged particles
for central collisions (Fig.~\ref{v23pt}), where the CMS experimental centrality bin $120\le N_{track}<150$ is compared to 
the hydrodynamic calculation with $N_{part}\ge 18$, corresponding to a similar centrality of $0-3\%$. We note that 
quantitative predictions of the hydrodynamic model depend on  the assumed initial shape and the parameters of the calculation 
\cite{Bozek:2011if,Bozek:2012gr,Bozek:2013uha,Bzdak:2013zma,Qin:2013bha}. 
Figure~\ref{figvn} shows the $p_T$ dependence of the elliptic flow coefficient $v_2$ of identified hadrons,
with a moderate but systematic mass scaling. The flow coefficients are 
obtained for events with $N_{part}\ge 14$, using the two-particle cumulant method and
the flow vector $Q$ defined by  the charged particles. In the experiment, a possible 
way to reduce the non-flow correlations would be to use  the peripheral centrality subtraction method 
\cite{ABELEV:2013wsa}, or to use the scalar product method, with the
 $Q$ vector defined in very forward pseudorapidity bins. In model calculations,
the cumulant and the scalar product methods give very similar results~\cite{Bozek:2013uha}. 
The experimentally observed mass splitting in p-Pb~\cite{ABELEV:2013wsa} is 
qualitatively similar as in the A-A collisions. Within hydrodynamics, the mass
splitting appears as a collective flow effect~\cite{Huovinen:2001cy}. 
The magnitude of the mass splitting in the p-Pb collisions can be reproduced
with a relatively high freeze-out temperature $T_f=150$~MeV, which suggest a relatively
short hadronic rescattering phase in the small systems, while a 
lower freeze-out temperature is used to reproduce the effect in A-A
collisions~\cite{Huovinen:2001cy}.

Figure~\ref{figv3} gives the prediction of the hydrodynamic model for the triangular flow coefficient of 
identified particles. 
The magnitude and the mass splitting of the triangular flow for identified hadrons
is much smaller than for the
elliptic flow (Fig.~\ref{figvn}); moreover, the ordering is inverted for small $p_\perp$. We have checked that 
the inverted mass ordering of $v_3$ in the range $p_\perp<500$~MeV comes as a result of resonance decays.

\begin{table}
\caption{\label{tablev2} $v_n\{2,|\Delta \eta| >2\}$, $n=2,3$, 
for all charged particles for various centrality classes, evaluated with the pseudorapidity gap $|\Delta \eta| >2$ and the cut 
$0.3 < p_T < 3$~GeV. The centrality is defined by $\langle N_{track} \rangle$ 
corrected 
as in \cite{Chatrchyan:2013eya}.}
\vspace{3mm}
\begin{tabular}{|l|ll|ll|}
\hline
$\langle N_{\rm trk} \rangle $  & \multicolumn{2}{|c|}{$v_2$} & \multicolumn{2}{|c|}{$v_3$} \\
   & model &  CMS \cite{Chatrchyan:2013eya} & model &  CMS \cite{Chatrchyan:2013eya} \\
\hline                      
154.5  &         0.057 &  0.063 & 0.023 & 0.02 \\              
96  &         0.059 &  0.057 & 0.022 & 0.016 \\ 
45  &         0.061 &  0.043 & 0.020 & 0.006 \\ \hline    
\end{tabular}

\end{table}

To assess the limits of the applicability of the hydrodynamic picture in p-Pb collisions, we compare the calculated  
integrated elliptic flow at three centralities to the CMS data (Table~\ref{tablev2}).
Clearly, as the systems becomes smaller the calculation deviates from the experimental values (especially for $v_3$), indicating that in peripheral 
p-Pb collisions the dissipative effects and the contribution of possible non-thermalized corona increase. 
Quantitative agreement of the model with the data on flow coefficients can be reached only for the most central interactions,
where the system is large enough to sustain a collective expansion phase that be described through relativistic viscous 
hydrodynamics.

We note that in small and short-lived systems part of the flow may be generated in the early pre-hydrodynamic phase~\cite{Chojnacki:2004ec,Gyulassy:2007zz}.
General arguments and numerical simulations show, however, that in many respects the pre-equilibrium flow leads to results similar to 
those of hydrodynamics extended to very early times~\cite{Broniowski:2008qk,Vredevoogd:2008id,Bozek:2010aj}. The p-Pb collisions offer a potential test ground to disentangle these effects.

In conclusion, we restate that the very fact that a strong mass hierarchy is seen in the highest-multiplicity 
p-Pb collisions at the LHC 
energies strongly suggests the collective nature of the evolution of the system. We have shown that the experimental 
results for the average transverse momentum 
and the elliptic flow coefficient $v_2$ can be described within the hydrodynamic approach, previously applied to this system.
 
While this paper was nearing completion, very similar qualitative conclusions were reached by Werner et al.~\cite{Werner:2013ipa}.

GT acknowledges the financial support received from the Helmholtz International
Centre for FAIR within the framework of the LOEWE program
(Landesoffensive zur Entwicklung Wissenschaftlich-\"Okonomischer
Exzellenz) launched by the State of Hesse, and from DOE under Grant No. DE-FG02-93ER40764. PB and WB acknowledge the support of the Polish
National Science Centre, grant DEC-2012/06/A/ST2/00390 and PL-Grid infrastructure.
One of us (GT) thanks Adrian Dumitru, Joern Putschke, Miklos Gyulassy, and Victoria Zhukova for constructive discussions.

\bibliography{hydr}

\end{document}